\documentclass[12pt,preprint]{aastex}

\begin{document}

\title{A Constant Bar Fraction out to Redshift $z\sim1$ in the
Advanced Camera for Surveys Field of the Tadpole
Galaxy}

\author{Bruce G. Elmegreen}
\affil{IBM Research Division, T.J. Watson Research Center, PO box
218,  Yorktown Heights, New York 10598 USA}
\email{bge@watson.ibm.com}

\author{Debra Meloy Elmegreen \affil{Vassar College,
Dept. of Physics \& Astronomy, Box 745, Poughkeepsie, NY 12604}}
\email{elmegreen@vassar.edu}

\and

\author{Amelia C. Hirst \affil{Vassar College,
Dept. of Physics \& Astronomy, Box 745, Poughkeepsie, NY 12604}}
\email{amhirst@vassar.edu}
\begin{abstract}
Bar-like structures were investigated in a sample of 186 disk
galaxies larger than 0.5 arcsec that are in the I-band image of
the Tadpole galaxy taken with the Hubble Space Telescope Advanced
Camera for Surveys. We found 22 clear cases of barred galaxies, 21
galaxies with small bars that appear primarily as isophotal twists
in a contour plot, and 11 cases of peculiar bars in clump-cluster
galaxies, which are face-on versions of chain galaxies. The latter
bars are probably young, as the galaxies contain only weak
interclump emission. Four of the clearly barred galaxies at
$z\sim0.8-1.2$ have grand design spirals. The bar fraction was
determined as a function of galaxy inclination and compared with
the analogous distribution in the local Universe. The bar fraction
was also determined as a function of galaxy angular size. These
distributions suggest that inclination and resolution effects
obscure nearly half of the bars in our sample. The bar fraction
was also determined as a function of redshift. We found a nearly
constant bar fraction of $0.23\pm0.03$ from $z\sim0$ to $z=1.1$.
When corrected for inclination and size effects, this fraction is
comparable to the bar fraction in the local Universe, $\sim0.4$,
which we tabulated for all bar and Hubble types in the Third
Reference Catalogue of Galaxies. The average major axis of a
barred galaxy in our sample is $\sim10$ kpc after correcting for
redshift with a $\Lambda$CDM cosmology. The average exponential
scale length is $\sim2$ kpc. These are half the sizes of local
barred galaxies and not likely to be influenced much by
cosmological dimming because the high-z galaxies are intrinsically
brighter. We conclude that galaxy bars were present in normal
abundance at least $\sim8$ Gy ago ($z\sim1$); the bars in clump
cluster galaxies may have formed from gaseous disk instabilities
and star formation rather than stellar disk instabilities, and bar
dissolution cannot be common during a Hubble time unless the bar
formation rate is comparable to the dissolution rate. If galaxy
interactions trigger bar formation more than bar destruction, then
the higher interaction rate in the past suggests relatively few
bars actually dissolved in a Hubble time.
\end{abstract}

\keywords{galaxies: evolution --- galaxies:  statistics --
galaxies: high-redshift -- galaxies: formation -- galaxies:
fundamental parameters}

\section{Introduction}

The fraction of barred galaxies as a function of redshift is an
important quantity for understanding galaxy evolution.  Models
predict that the secular evolution of bars may lead to dissolution
and bulge formation following the accretion of a sufficiently
large gas mass to the center (e.g., Combes \& Sanders 1981; Hasan
\& Norman 1990; Pfenniger \& Norman 1990; Bournaud \& Combes 2002;
Debattista et al. 2004). Recent 3-D chemodynamical models by
Immeli et al. (2004a) show that disk evolution depends on the
relative cooling rate: rapid cooling leads to early gas
instabilities, big star-formation complexes, and the mergers of
these complexes into a bulge, as in Noguchi (1999).  This phase
resembles high redshift chain galaxies if viewed edge-on (Cowie,
Hu, \& Songaila 1995; O'Neil et al. 2000; Elmegreen, Elmegreen, \&
Sheets 2004, hereafter Paper I; Immeli et al. 2004b), and what we
termed clump-clusters if viewed face-on (Elmegreen, Elmegreen, \&
Hirst 2004; hereafter Paper II). Slow cooling in other models by
Immeli et al. (2004a) leads to smoother stellar disks that may
become bar unstable if the effective Toomre-Q parameter drops to a
low value in the center. The possibility that bars also form in
clumpy galaxies was not discussed, nor was the formation of clumpy
bars. Little observational evidence has been presented for the
actual formation process of bars at high z.

Barred galaxies at intermediate and high redshifts are difficult
to find (van den Bergh et al. 1996, 2000, 2001; Abraham et al.
1999). The fraction of barred galaxies at $z>0.5$ has appeared to
be less than the local fraction (Abraham et al. 1999, Jogee et al.
2003), which van den Bergh et al. (2002) argued is not entirely a
selection effect. Nonetheless, because of the high angular
resolution required to find bars, and because there are inadequate
morphological clues in the rest-wavelength ultraviolet bands where
even low-redshift bars are indistinct (e.g., Windhorst et al.
2002; Sheth et al. 2003), it is possible that the fraction is
higher than these surveys indicate. Sheth et al. (2003) found four
barred galaxies with $z>0.7$ in the Near-Infrared Camera and
Multi-Object Spectrometer (NICMOS) observations of the Hubble Deep
Field-North (HDF-N), and commented that the corresponding fraction
of local galaxies with equally large bars ($\sim12$ kpc length)
was about the same.

Here we use the Hubble Space Telescope Advanced Camera for Surveys
(HST ACS) image of the Tadpole galaxy field (Tran et al. 2003) to
study bar fraction as a function of galaxy size, inclination, and
redshift using our sample of 186 spiral galaxies with major axes
greater than 10 pixels (see Paper I). Bars are identified by
morphology based on grayscale images and contour plots, and by
ellipticity and position angle twists in ellipse fits. Photometric
redshifts are from Benitez et al. (2004).

The Tadpole field observations have higher resolution than the
WFPC2 or NICMOS images of the HDF-N (Williams et al. 1996) and
HDF-S (Casertano et al. 2000) fields and they have about the same
depth at I band, so small galaxy features like bars at high z are
better seen in the Tadpole field. For the Tadpole field, the pixel
size is 0.05 arcsec and the stellar FWHM is 0.105 arcsec (Benitez
et al. 2004), whereas the HDF images have twice the pixel size
with WFPC2 and a stellar FWHM of 0.14 arcsec. All three fields
have about the same depth at $10\sigma$ in I band: 27.2 mag for
the Tadpole field (Benitez et al. 2004), 27.6 mag for the HDF-N
(Williams et al. 1996), and 27.7 mag for the HDF-S (Casertano et
al. 2000). The point spread function of NICMOS is twice that of
the WFPC2 (Sheth et al. 2003).

The present survey of 186 spiral galaxies in the Tadpole field
presents 22 cases of bars that are clearly visible in the
grayscale images, contour plots, and ellipse fits, and an
additional 21 cases where the contour plots show elongated
structures in the center that are twisted with respect to the
outer isophotes. The latter cases tend to be more poorly resolved
than the former, and may also be bars.

Triaxial bulges formed by galaxy mergers or bar dissolution (e.g.,
Friedli \& Martinet 1993; Friedli et al. 1996; Pfenniger 1999)
should be composed of a relatively smooth population of stars.
Evidence for significant clumpy structure in moderate-redshift
bars or evidence for asymmetric and irregular bars in
clump-cluster galaxies would suggest bar and bulge formation
processes operating mostly in the gas. This follows from the
observation of relatively large sizes and small numbers of clumps
in clump-cluster galaxies, which imply that the gas represents a
high fraction of the disk mass (Papers I and II).  Thus, bars in
clump-cluster galaxies may have formed from gaseous instabilities
like the clumps, rather than from instabilities in existing
stellar disks, which is the standard model of bar formation
(Sellwood \& Wilkinson 1993). To check this, we searched for
bar-like elongations in clump clusters and for clumpy bars in disk
galaxies with exponential profiles. Eleven examples of barred
clump-cluster galaxies and 2 examples of clumpy bars in
exponential disks were found.

The sample of high-redshift barred and inner twist galaxies is
large enough to allow corrections for selection biases resulting
from inclination and galaxy size.  It also begins to give the
distribution of bar fraction with redshift. The largest redshift
for a clearly barred galaxy in this survey is z=1.22 and the
largest for a galaxy with an inner isophotal twist is 1.84.

\section{Data and Analysis}

We used deep images of UGC 10214 in I band (F814W filter, total
exposure time per pixel 8180 s), which were obtained in April 2002
with the HST ACS as part of the Early Release Observations
proposal 8992 (Tran et al. 2003). The combined image is 3.9 arcmin
x 4.2 arcmin, with 0.05 arcsec pixels. The fully reduced image was
kindly made available by the ACS team in advance of publication,
using the methods of Blakeslee et al. (2003) and Tran et al.
(2003) but with more up-to-date reference files and a new damped
sinc ('lan\c{c}zos3') drizzling kernel for improved resolution
(Blakeslee, private communication). We determined photometric
parameters using the conversion $m = -2.5 \log({\rm counts}/{\rm
exposure\; time}) + {\rm zeropoint}$, with a zeropoint of 25.921
for I band (Blakeslee et al. 2003).

In Papers I and II, we classified galaxies according to their
morphology based on visual inspection of the V and I-band images
coupled with radial profiles and contour plots. In this paper, we
focus on the galaxies that appear to be normal spirals or clump
clusters.

Contour plots were made of each galaxy. The axial ratios of the
objects were determined by measuring lengths and widths from
contour plots at a contour level $2\sigma$, corresponding to a
surface brightness of 24.75 mag arcsec$^{-2}$. Ellipse fits were
made using the IRAF task {\it ellipse}, in which we specified the
central coordinate and the approximate ellipticity and position
angle, and stepped 1 pixel in radius. Not all galaxies could be
well fit by the ellipse task; 36 of the 186 galaxies in our
exponential disk sample failed to converge. We examined the
contour plots and images in order to provide a check on the
parameters obtained from the ellipse fits.

Galaxies were considered to be clearly barred if they looked
barred on grayscale images and contour plots, and also had a
maximum ellipticity accompanied by a change in position angle at
the same radius.  Galaxies with poor pixel resolution often did
not have clear bars on gray scale images, but if they still had
contours that twisted in the centers, as in clearly barred
galaxies, we considered them to be bar-like anyway. We refer to
this latter type as galaxies with inner isophotal twists.

\section{Results}

\subsection{Barred and Inner Twist Galaxies in the Tadpole Deep
Field}

Barred and inner-twist galaxies found in the deep Tadpole Galaxy
field are listed in Table 1.  The galaxy number and redshift are
from Benitez et al. (2004); the bar type, size, exponential scale
length, ratio of axes, and apparent I-band magnitude are from our
measurements. The apparent I-band magnitude is measured down to
$1\sigma$ of the noise, which is $25.5$ mag arcsec$^{2}$. The
absolute magnitude comes from the distance modulus, which is given
by the equation (Carroll, Press, \& Turner 1992)
\begin{equation}
DM\left(z_{gal}\right)=25+5\log\left({{c\left(1+z\right)}\over
{H_0}}\int_0^{z_{gal}}
\left[\left(1+z\right)^2\left(1+\Omega_Mz\right)-z\left(2+z\right)
\Omega_\Lambda\right]^{-1/2}dz\right)\label{eq:dm}
\end{equation}
We assume the cosmology determined from WMAP (Spergel et al.
2003): $\Omega_\Lambda=0.732$, $\Omega_M=0.268$, and $H_0=71$ km
s$^{-1}$ Mpc$^{-1}$.  Entries at the bottom of the table were not
catalogued by Benitez et al. (2004) because they are close to the
Tadpole Galaxy or near a bright star. Instead of giving a Benitez
et al. catalog number for these, we give in the table notes the
coordinates in RA and DEC for epoch 2000.

\begin{figure}
\plotone{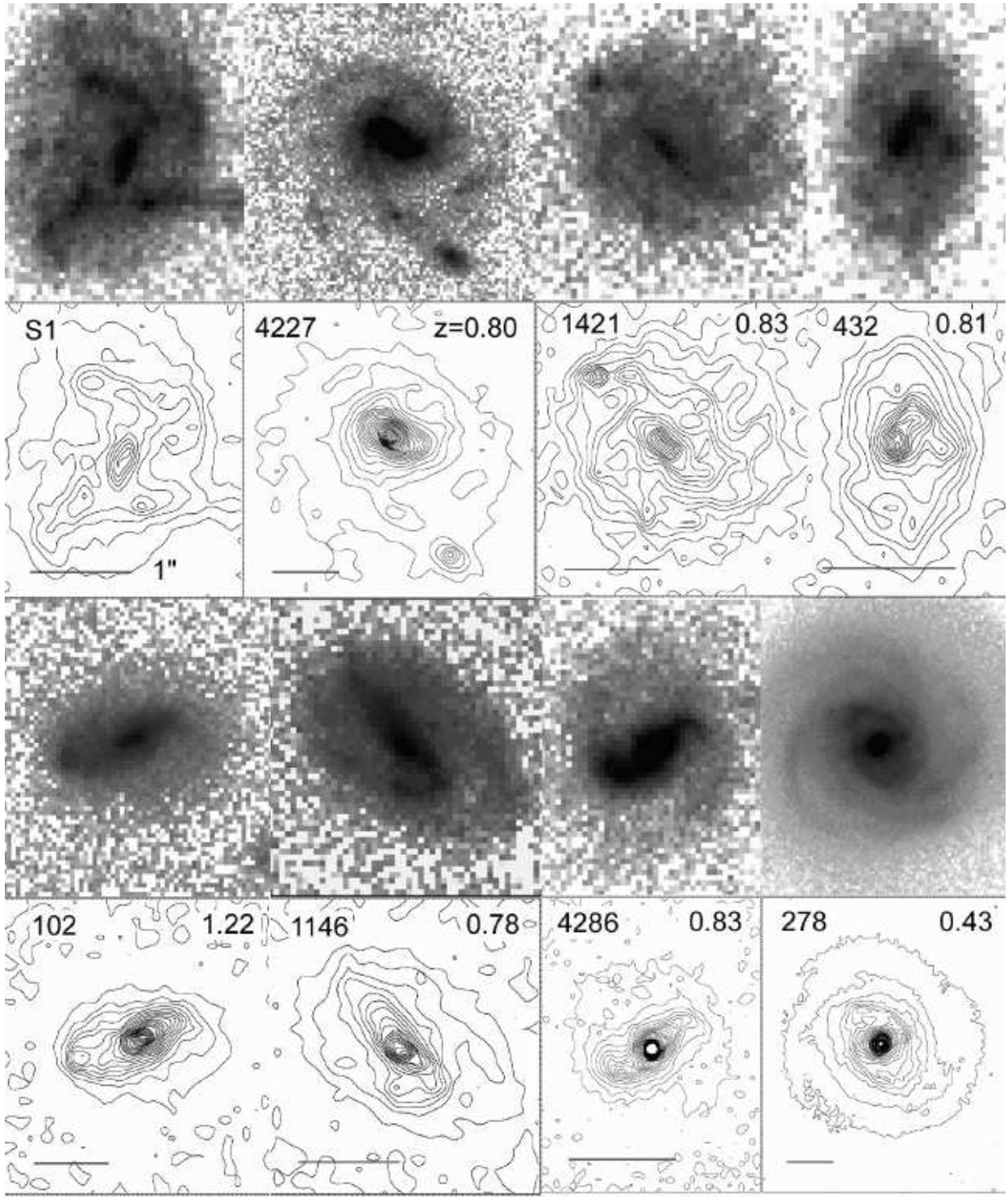} \caption{I-band images of 8 barred galaxies,
showing logarithmic grayscale on top and linear contour plots on
the bottom. Arc second lengths are indicated by lines on the
contour plots. The galaxies numbers from Benitez et al. (2004) are
on the top left of each panel, and the redshifts are on the top
right, when available.}\label{fig:8bars}\end{figure}

Figure \ref{fig:8bars} shows a sample of I-band images of 8 barred
galaxies in logarithmic grayscale and linear contour plots.  Most
look like normal galaxies in the nearby Universe, having spirals
outside the bars, some of which are grand-design spirals, and
end-of-bar enhancements to star formation. The galaxies are named
by Benitez et al. number (except for S1), and the redshifts and
scales are indicated. The nicest grand design spirals in this
figure are numbers 4227, 102, 1146, and 4286, with redshifts
between 0.78 and 1.22. van den Bergh et al. (2000) noted that
grand design galaxies are rare beyond $z=0.3$, although they
showed one at $z=0.9$.  The bars in our survey also extend to
higher $z$ than the limit of $z\sim0.5$ found by van den Bergh et
al. (2000).

\begin{figure}
\plotone{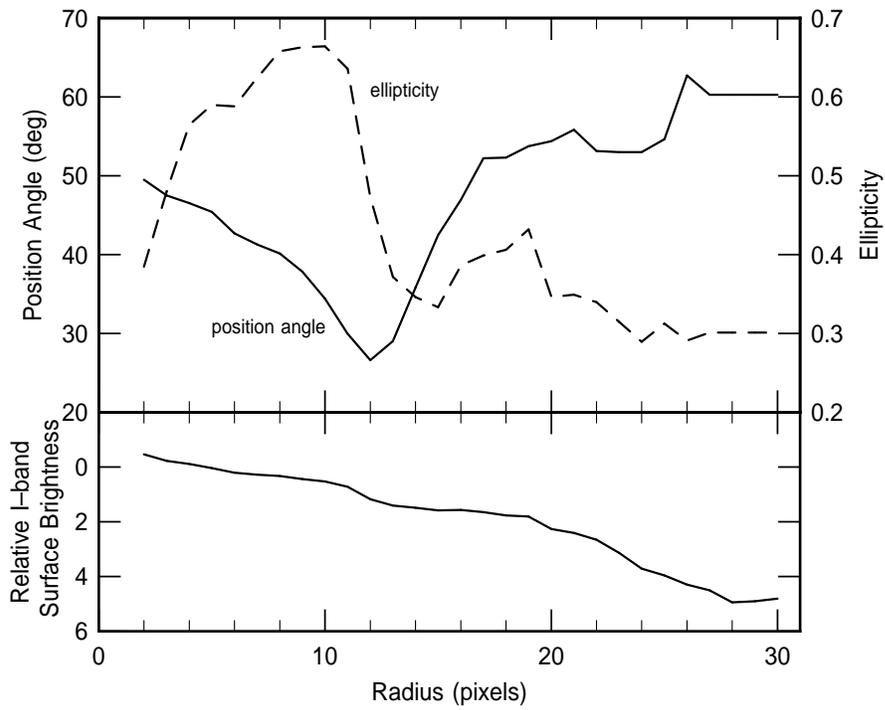} \caption{(top)  Position angle and ellipticity
for galaxy 1146. The bar ends at 11 pixels, where the ellipticity
drops and the position angle changes. (bottom) Radial profile of
the average relative surface brightness in I band based on the
ellipse fits.}\label{fig:radial}
\end{figure}

Figure \ref{fig:radial} illustrates ellipse and position angle
fits to the light distribution in galaxy 1146, not corrected for
projection. The end of the bar at 11 pixels is evident from the
change in the position angle and decrease in ellipticity as the
bar ends and the spirals begin beyond this radius. Figure
\ref{fig:radial} also shows the average disk light profile,
obtained from the ellipse fits; it is approximately exponential,
as is the light profile along the bar (not shown). The maximum
projected ellipticities for our barred galaxies are between 0.35
and 0.72.

\begin{figure}
\plotone{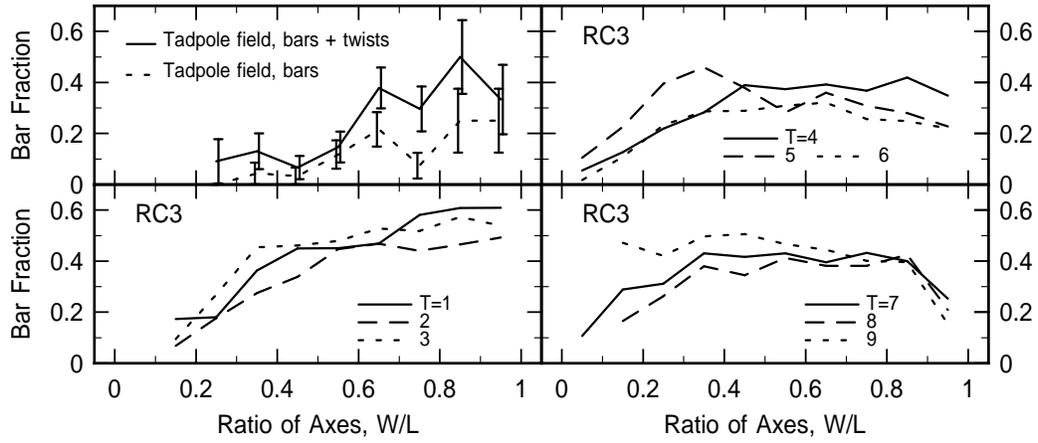} \caption{Bar fraction as a function of the ratio
of the minor to the major axis for galaxies in our sample, the
Tadpole deep field (top left), and for SB galaxies in the RC3,
sorted by T-type. Statistical errors are shown for the Tadpole
field fractions.  Errors for the RC3 points are $\pm0.04$.}
\label{fig:barfractwl}
\end{figure}

Figure \ref{fig:barfractwl} shows the effects of inclination on
bar recognition for galaxies in our survey (top left) and for all
of the spiral galaxies in the Third Reference Catalogue of Bright
Galaxies (RC3, de Vaucouleurs et al. 1991). The plotted quantity
for the RC3 survey is the fraction of all spiral galaxies with T
types 1 through 9 ($=1$ for Sa, 2 for Sab, etc.) that were also
classified as type SB. Each curve is for a different T type.
Statistical errors determined from
$\left(f\left[1-f\right]/N\right)^{1/2}$ for fraction $f$ and
total number $N$ are all around $\pm0.04$ for RC3 galaxies and are
not shown. Many of the RC3 galaxies with T types have no bar type
classification (SA, SAB, or SB) so the plotted fraction represents
a lower limit to the bar fraction. The decrease in bar fraction
for decreasing ratio of minor to major axis, $W/L$, indicates that
bars are hard to recognize in edge-on galaxies. The same
difficulty apparently arises for the Tadpole field, where
statistical errors, determined in the same way as for the RC3
sample, are shown.  Errors in measurement of either $W$ or $L$ are
approximately one pixel, so $W/L$ is accurate to better than 10\%
for typical inclinations and sizes.  Surface brightness dimming
with z reduces $W$ and $L$ together if the ellipticity on a
contour diagram is independent of radius.  For barred galaxies,
the ellipticity strongly depends on radius in the inner regions
but varies weakly in the outer regions where we measure $W/L$.

A random distribution of thin circular disks would have a uniform
distribution of $W/L$ down to the ratio of the intrinsic disk
thickness to the diameter, with a slight bump at this lower limit
before it drops to zero (Sandage, Freeman, \& Stokes 1970; Paper
II).  If there are in fact equal numbers of galaxies in equal
intervals of $W/L$, and the actual presence of a bar is
independent of our perspective, then the gap between the curve and
the asymptote at high $W/L$ is the result of unrecognized bars. We
calculate from the RC3 plots that between $\sim0$\% of the late
type galaxies and $\sim20$\% of the early type galaxies have bars
that are missing because of high inclinations.  The asymptotic SB
fractions are taken from the averages for $W/L>0.5$ and are
plotted in Figure \ref{fig:barfracttype} below.

Based on the upper left panel in Figure \ref{fig:barfractwl},
$25$\%--50\% of the galaxies in the Tadpole field have missing
bars because of high inclinations. These limits come from the
gradual rise in the bar fraction with $W/L$: If we consider the
average barred fraction for the most face-on half, with $W/L>0.5$,
to be representative of the intrinsic barred fraction, and use the
relative drop in the average for $W/L<0.5$ as an indication of the
missing barred fraction, then we get 25\%. If we consider the peak
barred fraction at $W/L\sim1$ to be indicative of the intrinsic
fraction, and consider everything below this fraction to be
missing, then we get 50\%.

Evidently, deep field bars are harder to recognize at low
inclination than nearby bars. This is probably because deep field
bars are closer to the detection limits imposed by pixel size and
surface brightness than local bars. The ratio of these two
missing-bar fractions should be the correction factor to the
relative number of bars in the Tadpole field compared to the local
field resulting from more sensitive inclination losses in the
Tadpole field.  This correction depends on which local Hubble type
is used for comparison.

The asymptotic barred galaxy fraction at high $W/L$ in the Tadpole
field is $\sim20$\% for the bars alone and $\sim40$\% for the
combined bars and inner twist galaxies. The asymptotic fractions
for RC3 galaxies of types SB (solid lines), SAB (dashed lines),
and SA (dotted lines) are plotted in Figure
\ref{fig:barfracttype}.  In each case, the bottom curve is the
fraction relative to the total number of galaxies having the same
spiral T type in the RC3, and the top curve is the fraction
relative to the total number having any SB, SAB, or SA type within
that T type (not all galaxies with spiral T types have classified
bar types).  There is either a bias in the RC3 or an intrinsic
difference in bar structures such that intermediate Hubble types
(T$=4,5,6$ corresponding to Sbc, Sc, and Scd) are more likely to
be classified SAB than SB compared to early and late Hubble types.
This same pattern of optical bar fraction versus T type was shown
by Eskridge et al. (2000) using a smaller sample of 186 RC3
galaxies. To make Figure \ref{fig:barfracttype}, we determined the
averages using only RC3 galaxies with $W/L>0.5$, i.e., using only
the most face-on half of the sample, for which the bar type is
most evident. There are 7134 spiral galaxies with classified T
types and $W/L>0.5$ in the RC3 and 12471 spiral galaxies with
classified T types and $W/L>0.1$. The bar fractions in Figure
\ref{fig:barfracttype} are slightly higher than those usually
estimated from the RC3 because we do not include the low
inclination galaxies, which have a higher probability that their
bars are not recognized.

Eskridge et al. (2000) obtained an SB fraction of 56\% and an SAB
fraction of 16\% averaged over all of the spiral T types and
inclinations in their sample of 186 galaxies taken in the H band.
van den Bergh et al. (2002) classified 101 local galaxies
according to the DDO system and obtained a local bar fraction
(equivalent to SB here) of 28\% in optical bands, again averaging
over spiral type and inclination. van den Bergh et al. (2002) also
blurred their galaxies to resemble $z=0.7$ galaxies observed with
0.04 arcsec resolution (comparable to the pixel size of the
Tadpole field, which is 0.05 arcsec). The SB fraction dropped to
19\% with this blurring, suggesting that $\sim30$\% of local bars
would be unrecognized at $z\sim0.7$.

\begin{figure}
\plotone{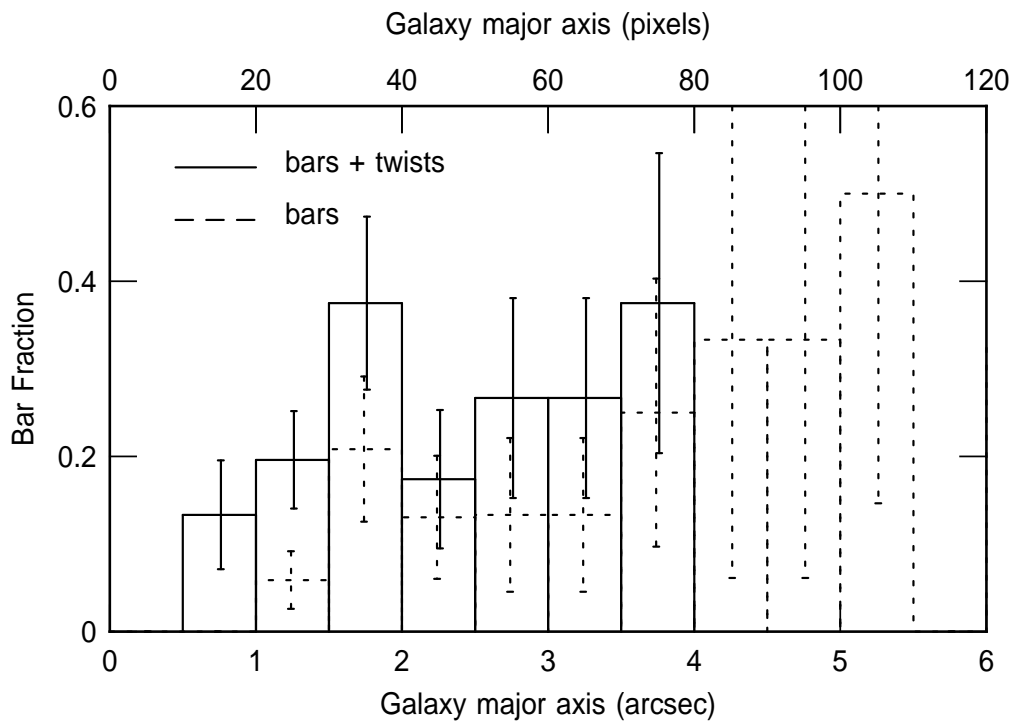}\caption{Bar fraction is shown as a function of
the angular size of the major axis for galaxies in the Tadpole
deep field, with statistical error bars indicated. The solid line
is the sum of the counts for galaxies with clear bars and
isophotal twist bars, whereas the dashed line is the count for
clear bars only. A clear bar is visible on both a grayscale image
and as an isophotal twist in the intensity contours. An isophotal
twist bar shows up primarily as an inner isophotal twist in the
contours. The galaxies with the largest angular sizes have clear
bars. The smallest galaxies with isophotal twists may also have
bars, but these bars are not obvious on the grayscale images
because they subtend only a few pixels.}\label{fig:barvsl}
\end{figure}

The distributions of angular size for the barred and inner twist
galaxies in the Tadpole field are shown in Figure
\ref{fig:barvsl}. The scale in arcsec is shown on the bottom axis
and the scale in pixels is on the top.  The bar fraction increases
with angular size because bars with more pixels are easier to
recognize. Also, the bars in galaxies with large angular sizes are
easily recognized as bars in both the grayscale images and in
contour plots. The counts for these clear bars are drawn as dashed
lines in the figure.  At the largest angular sizes, all of the
bars are clear bars.  The counts for galaxies whose bars could
only be recognized as twisted inner isophotes on contour plots are
added to the clear-bar counts in the figure, and this sum is drawn
as a solid line.  Galaxies with small angular sizes tend to show
only the inner twisted isophotes without any clear bar on the
grayscale images, so the left-hand part of the figure is dominated
by these solid lines.

\begin{figure}
\plotone{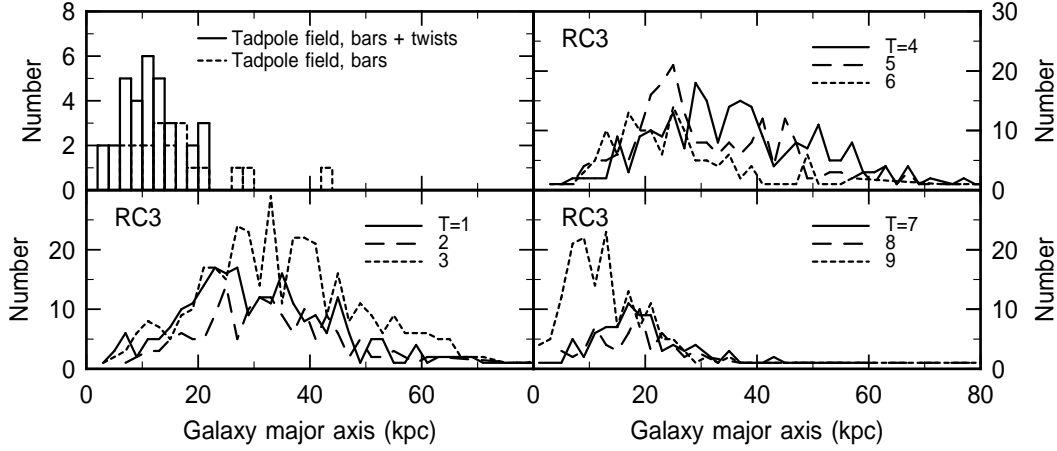} \caption{Histograms of the numbers of galaxies as
a function of major axis size for the Tadpole field and for the
RC3, where they are separated by spiral T type. The physical sizes
for the Tadpole field galaxies are derived from the angular sizes
and the redshift using Eq. 2. The Tadpole field barred galaxies
appear smaller than most local barred galaxies, by a factor of
$\sim2$, making them comparable to the smallest local galaxies, of
type SBm.} \label{fig:barfractsize}
\end{figure}

A histogram of absolute galaxy diameter for the Tadpole field is
shown in the top left of Figure \ref{fig:barfractsize}. The
conversion from angular size $\theta$ and galaxy redshift
$z_{gal}$ to physical size $L$ was made from the equation
(Carroll, Press, \& Turner 1992)
\begin{equation}
L={{c\theta}\over{H_0\left(1+z\right)}}\int_0^{z_{gal}}
\left[\left(1+z\right)^2\left(1+\Omega_Mz\right)-z\left(2+z\right)
\Omega_\Lambda\right]^{-1/2}dz\label{eq:l}
\end{equation}
The diameter is taken to be the major axis of the second isophotal
contour, which is $2\sigma$ above zero for rms noise $\sigma$. The
other panels in Figure \ref{fig:barfractsize} are for all SB type
galaxies in the RC3, using the radial velocities with $H_0=71$ km
s$^{-1}$ Mpc$^{-1}$ and the corrected face-on diameters. Figure
\ref{fig:barfractsize} indicates that most of the galaxies in the
Tadpole field are about half the size of the local galaxies with
early and intermediate Hubble types.  The Tadpole field galaxies
have sizes comparable to the latest-type local galaxies, $T=9$ for
type SBm. There is a tendency for the clearest bars in the Tadpole
field (dotted line) to be in the largest galaxies. Small galaxies
in the Tadpole field may also have bars, but they show up in our
data only as inner isophotal twists (pure solid lines).

The small sizes of the high-z galaxies could be the result of
surface brightness dimming, but the actual values of the surface
brightnesses suggest this is not the case. Equations \ref{eq:dm}
and \ref{eq:l} combine to predict surface brightness dimming
varies as $10\log\left(1+z\right)$ in mag arcsec$^{-2}$, which is
the standard result (e.g., Lubin \& Sandage 2001).  For $z$
between 0.5 and 1.2 as in this sample, the dimming is between 1.8
and 3.4 mag arcsec$^{-2}$. Each magnitude of dimming corresponds
to a loss of approximately one exponential scale length from the
outer part of a galaxy, and since typical galaxy disks in the
local Universe have $\sim4$ scale lengths out to 25 mag
arcsec$^{-2}$, a dimming of 2 mag arcsec$^{-2}$ would cut the
galaxy in half. However, Figure \ref{fig:radial} shows a $z=0.78$
galaxy that has more than 4 exponential scale lengths, as do many
others in our survey (see next paragraph). In addition, the 2
$\sigma$ contour limit used for the galaxy size corresponds to
24.75 mag arcsec$^{-2}$ in I band, which is comparable to B or V
bands for more local galaxies where the radius at 25th mag
arcsec$^{-2}$ is measured.  Corrected for dimming, the restframe
outer contour at high $z$ would be $\sim23$ mag arcsec$^{-2}$ or
brighter in most cases, suggesting that the high-z galaxies in our
survey are selectively brighter than local galaxies. As this
restframe brightening offsets the cosmological dimming, the
physical sizes of the Tadpole-field barred galaxies in our study
really are small. This is consistent with other studies that show
surface brightness evolution for disk galaxies with z-dependent
restframe properties (Bouwens \& Silk 2002). It differs from
studies (Simard et al. 1999; Ravindranath et al. 2004) that do not
show surface brightness or size evolution, but in those cases the
galaxies are selected to have the same restframe properties. Our
sample is too small to make this selection; i.e. we do not see a
population of high-z barred galaxies that have restframe surface
brightnesses comparable to those of local barred galaxies.  There
is no way to tell from this survey if such a barred population is
present, and, if it is, whether it has the same proportion to
non-barred galaxies as the high surface-brightness population
studied here.

\begin{figure}
\plotone{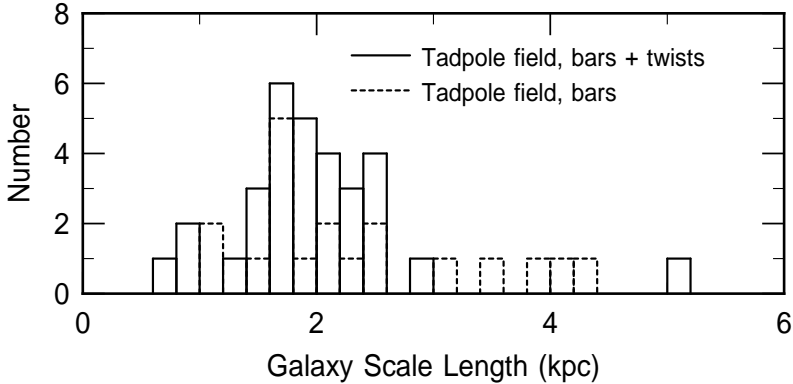} \caption{The distribution of exponential scale
lengths corrected for redshift according to equation 2. The scale
lengths are not affected by cosmological surface brightness
dimming.  Like the absolute sizes of Tadpole field galaxies, the
average scale length is smaller than the scale lengths in local
barred galaxies.}\label{fig:scal}
\end{figure}

To check the sizes of high-z barred galaxies another way, we
measured the exponential scale lengths for each of the 22 clear
bars and 21 inner twist galaxies using ellipse fits in the I band.
The scale lengths are independent of dimming. The results are in
Table 1 and a histogram is in Figure \ref{fig:scal}. The average
scale length is $\sim2$ kpc, comparable to that found in recent
simulations of disk galaxy formation (Robertson et al. 2004), but
smaller than in the Milky Way and other local galaxies by a factor
of $\sim2$. The average ratio of the major axis to the scale
length for the Tadpole field barred galaxies is 7.2, and the
average ratio for the inner twist galaxies is 5.6. These ratios
are normal compared with local galaxies, considering a Freeman
central surface brightness of 21.6 mag arcsec$^{-2}$ in B band and
an outer radius at 25 mag arcsec$^{-2}$ in B band. The local
numbers give an average of $\sim3.3$ scale lengths in the disk and
a corresponding ratio of major axis to scale length equal to
$\sim6.6$.  The similarity between the numbers of scale lengths in
the local and Tadpole field disks implies the distant galaxies are
not truncated significantly by surface brightness limits.  Thus,
Figures \ref{fig:barfractsize} and \ref{fig:scal} both imply the
Tadpole field galaxies are intrinsically small.

\begin{figure}
\plotone{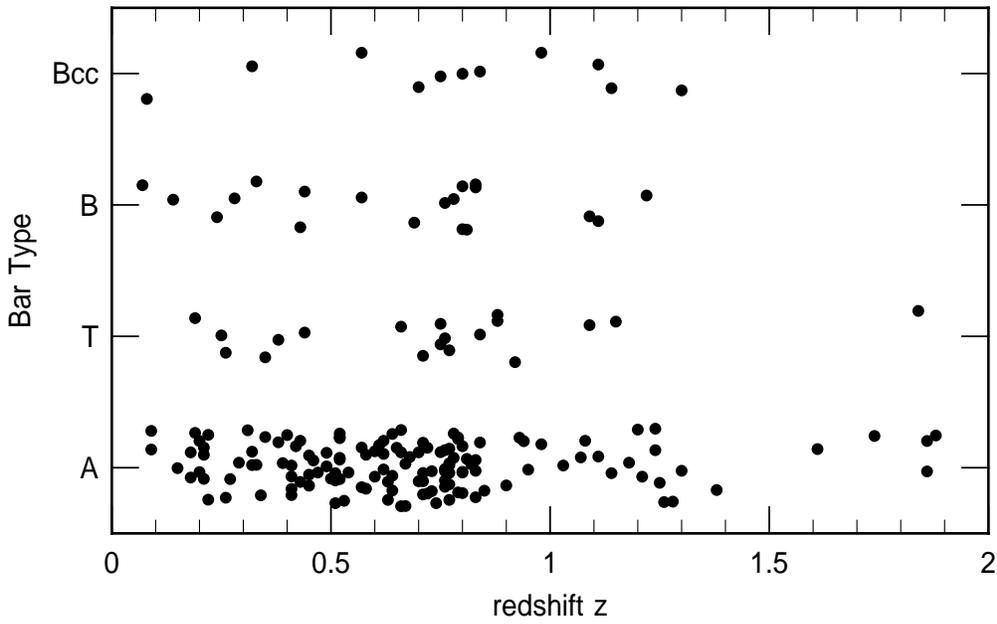} \caption{The distribution of redshifts is shown
for galaxies with bars (B), isophotal twists (T), no bars (A), and
clump-cluster galaxies with bars (Bcc). They all span similar
redshift ranges, although bars are absent beyond z=1.3, possibly
from our inability to resolve them and the usual obscurity of bars
in the restframe uv.}\label{fig:barvsz}
\end{figure}

\begin{figure}
\plotone{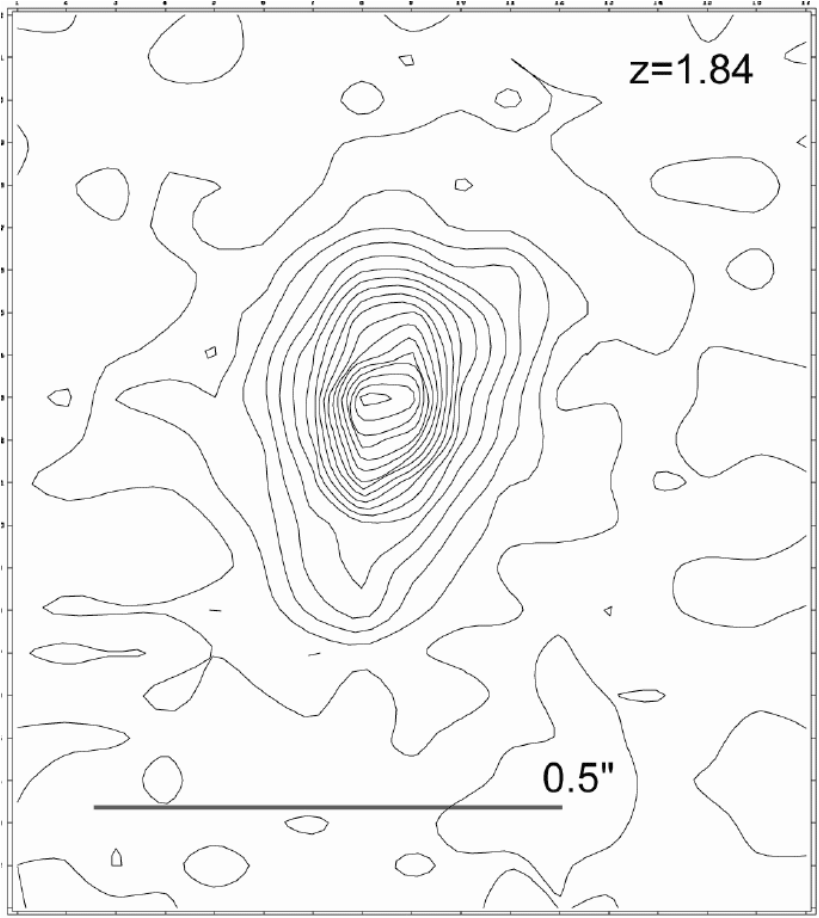} \caption{The highest redshift galaxy that might
have a bar is number 3555 in Benitez et al. (2004), shown here.
The bar is not clear because it is small, but a twist in the inner
isophotes is present nevertheless, suggesting the presence of a
small bar.}\label{fig:3555}
\end{figure}

\begin{figure}
\plotone{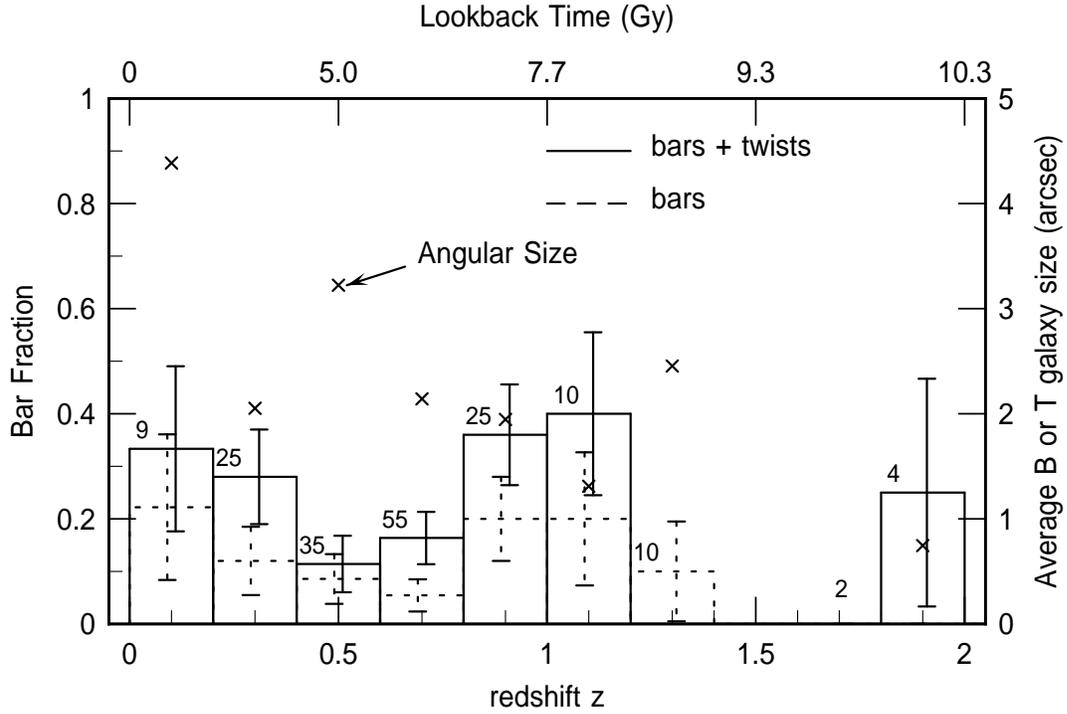} \caption{The bar fraction is shown as a function
of redshift for bars and twists (solid line) and for bars only
(dashed line). The x's indicate the average angular sizes of
galaxies in each redshift bin, using the axis on the right.
Statistical errors are shown, along with the total number of
galaxies in each bin. The corresponding lookback time for each z
is shown along the top axis.  The average bar fraction out to
$z=1.1$ is $0.23\pm0.03$.  This fraction should be increased by a
factor of $\sim2$ to account for inclination and resolution
effects.}\label{fig:barfractz}
\end{figure}

The distributions of the disk galaxy types: non-barred (A), inner
twist (T), barred (B), and barred clump-cluster galaxies (Bcc),
are shown as a function of redshift in Figure \ref{fig:barvsz}.
The points on the ordinate are given a random offset to avoid
overlaps. The bars seem to stop at $z\sim1.3$, but this is
probably because the ability to resolve and recognize distant bars
diminishes. For example, the galaxy at $z=1.22$ in Figure
\ref{fig:8bars} clearly shows a bar even in this restframe blue to
uv passband, but the bar is only $\sim7$ pixels long and beginning
to get lost in the pixel noise.  Galaxies with inner isophotal
twists continue up to $z=1.84$, and this galaxy, which is number
3555 in Benitez et al (2004), is shown in Figure \ref{fig:3555}.
The greater $z$ limit for twist galaxies is consistent with
Figures \ref{fig:barvsl} and \ref{fig:barfractsize}, where the
smallest angular-size galaxies show only isophotal twists without
obvious bars. There is a good correlation between angular size and
$z$ in our data in the sense that high $z$ galaxies have
systematically smaller angular sizes (see crosses in Fig.
\ref{fig:barfractz}).

The bar fraction is shown as a function of redshift $z$ in Figure
\ref{fig:barfractz}.  The average major-axis galaxy angular size
in arcsec is plotted for each redshift bin as a cross. The numbers
above the bins are the total numbers of galaxies.  The bar
fraction is approximately constant with redshift out to $z\sim1$,
having values between 0.2 and 0.4. Corrections for inclination
could double this fraction (see Fig. \ref{fig:barfractwl}).

Lookback time is given on the top axis of Figure
\ref{fig:barfractz}.  This is calculated from the equation
(Carroll, Press, \& Turner 1992)
\begin{equation}
t_{\rm lookback}= {{1}\over{H_0}}\int_0^{z_{gal}}
\left(1+z\right)^{-1}
\left[\left(1+z\right)^2\left(1+\Omega_Mz\right)-z\left(2+z\right)
\Omega_\Lambda\right]^{-1/2}dz
\end{equation}
Lookback time is useful for estimating whether the clumps in these
galaxies are old enough to make bulges if they coalesce, and for
assessing the epoch of bar formation (see discussion section).

\subsection{Clumpy-Bar Galaxies and Barred Clump-Cluster Galaxies}

In our survey of barred galaxies in the Tadpole field, we noticed
several galaxies with exponential disks and very clumpy bars, and
several others with purely clumpy disks (called clump clusters in
Paper II) and bar-like objects in those disks. In this paper, we
have called these two odd cases clumpy-bar galaxies and barred
clump-cluster galaxies, respectively. The clumpy-bar galaxies are
from the sample of 186 exponential disk galaxies, whereas the
barred clump-cluster galaxies are from the sample of 86
clump-cluster galaxies, which do not have exponential disks, as
studied in Paper II.

\begin{figure}
\plotone{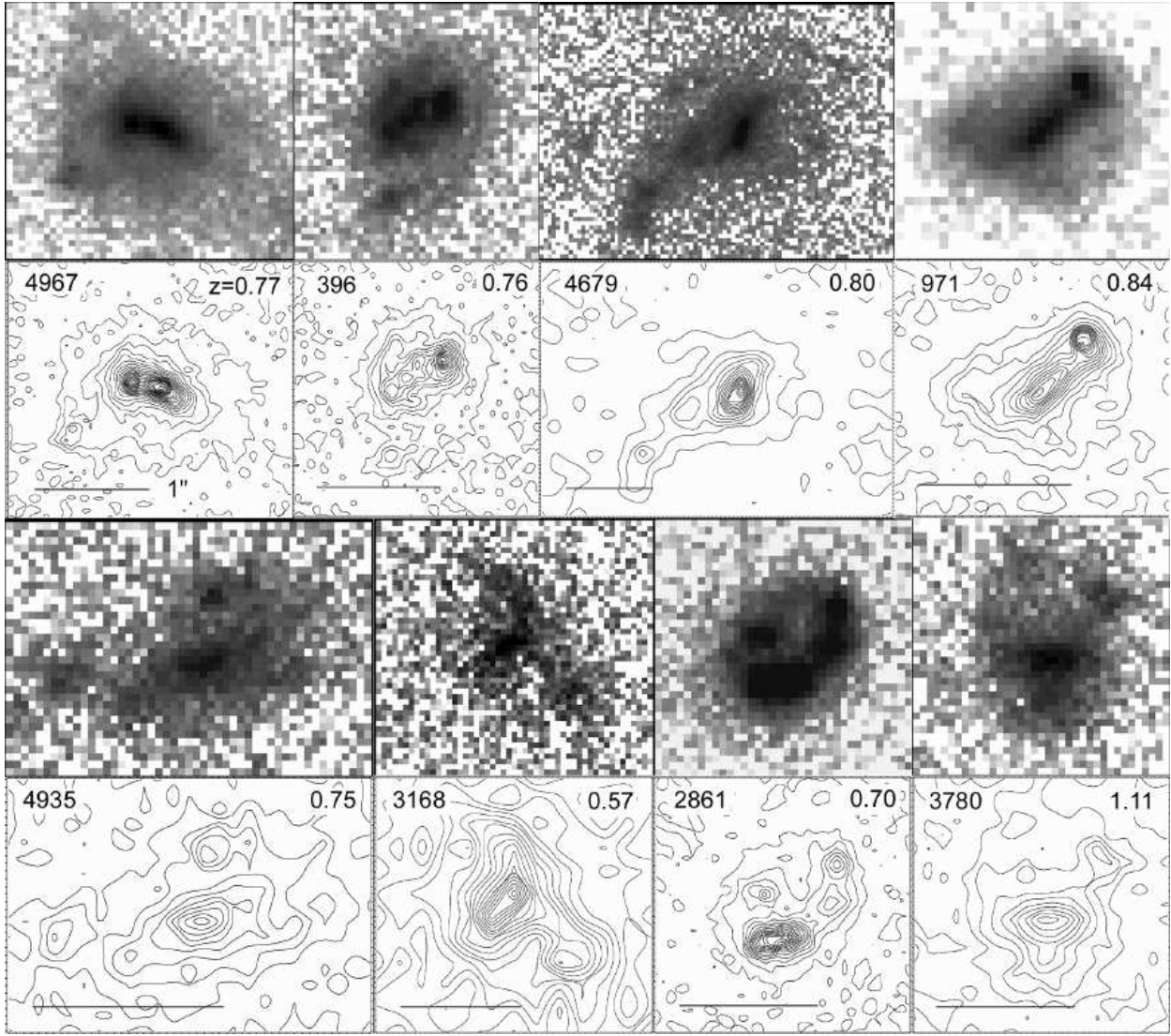} \caption{I-band logarithmic grayscale and linear
contour plots are shown for spiral galaxies with clumpy bars
(numbers 4967 and 396) and barred clump-cluster galaxies. The
redshifts and scales for each galaxy are indicated.}
\label{fig:8oddbars}
\end{figure}

Figure \ref{fig:8oddbars} shows 8 examples of these unusual cases.
Table 1 gives the properties of the two clumpy-bar galaxies
(numbers 4967 and 396), and Table 2 has the 11 barred
clump-cluster galaxies.

\begin{figure}
\plotone{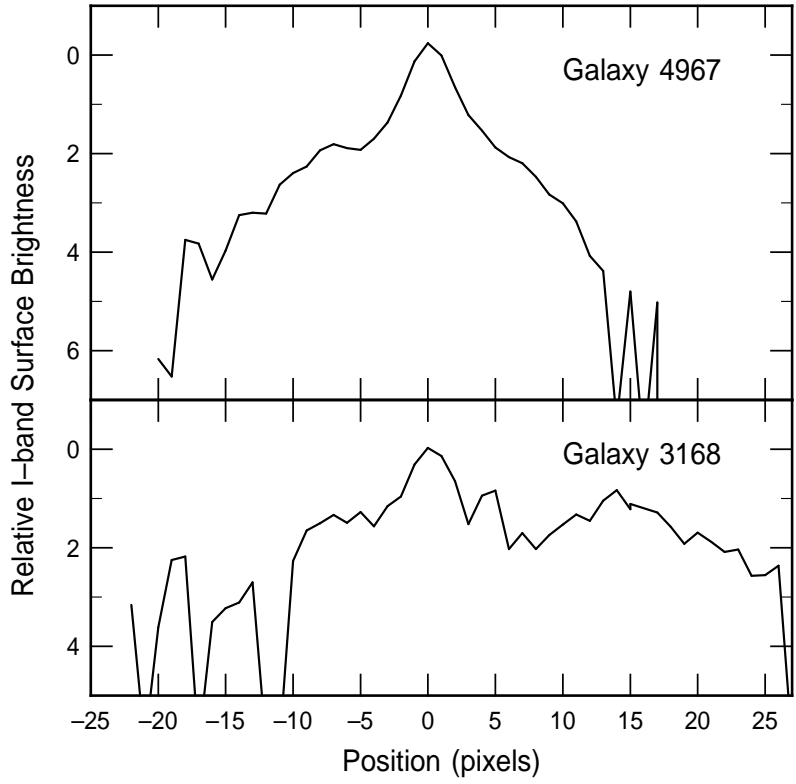} \caption{Radial profiles based on major axis
cuts are shown for galaxy 4967, which has an exponential disk and
a clumpy bar, and galaxy 3168, a clump-cluster galaxy with a
clumpy bar and no exponential disk.}\label{fig:radial147310}
\end{figure}

Figure \ref{fig:radial147310} shows the relative surface
brightness profiles in two of these galaxies using the task {\it
pvector} in IRAF for a strip 3 pixels wide across the galaxy
image.  The orientation of each strip was approximately along the
major axis. Galaxy 4967 is a clumpy-bar galaxy and is shown at the
top left in Figure \ref{fig:8oddbars}. Galaxy 3168 is a barred
clump-cluster galaxy and is shown as the second image from the
left on the bottom of Figure \ref{fig:8oddbars}. The
exponential-like profile in galaxy 4967 is evident, as is the lack
of a similar profile in galaxy 3168.  This lack of an exponential
profile defines the clump-cluster and chain galaxy types (Papers I
and II).

\section{Discussion}

The results in the previous section suggest there are normal
barred galaxies up to at least a redshift of $z\sim1$.  There are
also bars or bar-like regions in exponential-disk galaxies that
are mostly composed of bright clumps, and there are bar-like
regions inside completely clumped galaxies, which we have called
clump-clusters.   Beyond $z\sim1$, bars become indistinct because
the galaxies have small angular sizes and the rest passband goes
into the ultraviolet where bars are not prominent (e.g., Sheth et
al. 2003).  Nevertheless, we still see galaxies out to $z=1.84$
that appear to contain bars in the form of isophotal twists in
their inner regions (Fig. \ref{fig:3555}).

The time interval corresponding to the histogram interval in
Figure \ref{fig:barfractz}, namely $\Delta z=0.2$, becomes
comparable to the formation time of a bar near $z=2$ (i.e.,
$\Delta t_{\rm lookback}=3.4\times10^8$ years). Thus the bar
fraction should begin to vary with $z$ significantly at this point
for physical reasons, and the bar morphology should vary too if
some bars are only part-way to formation.  Galaxies that form
disks late push back this epoch of bar irregularity to lower $z$.
Our clumpy bars and barred clump-clusters could be in this
category.

The bar fraction between $z=0$ and $z=1.2$ varies between 0.2 and
0.4. The average fraction for clear bars determined from the total
number of them (18) divided by the total number of disk galaxies
out to $z=1.2$ (159) is $0.11\pm0.03$. The average fraction for
galaxies whose only evidence for a bar is an inner twisted
isophote is the same because the number of these cases is also 18.
The average fraction for both together is $0.23\pm0.03$.

These bar or twist fractions are smaller than the local SB
fraction from Figure \ref{fig:barfracttype}, which ranges from
$\sim0.5$ for early type galaxies to $\sim0.35$ for intermediate
to late Hubble types.  The Tadpole field bar fraction requires
upward corrections, however. We suggested on the basis of Figure
\ref{fig:barvsl} that the smallest galaxies in our sample could
have bars that are not recognized, whereas this is not a problem
in the local sample. We also suggested on the basis of Figure
\ref{fig:barfractwl} that the loss of bars to inclination effects
is greater in the Tadpole field than for local galaxies. Thus our
recognition fraction for bars in the deep field is less than for
local bars, perhaps by a factor of 2.  For the least inclined
sub-sample in Figure \ref{fig:barfractwl}, and the largest angle
subsample in Figure \ref{fig:barvsl}, the bar plus twist fraction
is $\sim0.4$, with most of the bar-like galaxies actually showing
their bars clearly in the grayscale images of the big or face-on
galaxies. This corrected bar fraction is about the same as the
local SB fraction from Figure \ref{fig:barfracttype}.

The small absolute sizes of the barred galaxies in Figures
\ref{fig:barfractsize} and \ref{fig:scal}, combined with the
apparently normal bar fractions at high $z$, seem to rule out
models in which disk galaxies continue to build up by mergers
after $z\sim1.5$. A merger will destroy a disk, so none of the
small barred disk galaxies studied here can turn into bigger disk
galaxies today after significant mergers. They either remain
small, grow by gradual, non-disruptive accretions, or turn into
ellipticals after a merger.

The nearly constant bar fraction with $z$ combined with the
theoretical (Noguchi 1987, Gerin, Combes \& Athanassoula 1990;
Berentzen et al. 2004) and observational (Elmegreen, Elmegreen, \&
Bellin 1990) evidence that close interactions (not mergers) form
bars, also suggests that most galactic bars have not been
destroyed over a Hubble time. If they did dissolve in
significantly less than a Hubble time, then the bar fraction
should be the result of an equilibrium between formation and
destruction rates. As the interaction rate was probably much
higher in the past than today, the bar formation rate was probably
higher too. Thus the bar fraction should have been higher in the
past if interactions were an important bar-formation process and
self-destruction followed formation in significantly less than a
Hubble time (as it does, for example, in models by Bournaud \&
Combes 2002). The constant bar fraction over time suggests there
has been relatively little bar destruction.

This conclusion differs from that of Block et al. (2002) and Das
et al. (2003) who suggest observations support the bar dissolution
theory.  Block et al. based this conclusion on the lack of local
galaxies with smooth stellar disks (the opposite of a lack of
bars, as might be expected) and noted that simulations require
significant gas accretion to keep stellar spirals present. Then
they concluded that such accretion could also rejuvenate bars {\it
if} they were previously dissolved, but did not offer any
evidence, aside from the usual simulation results, that real bars
have dissolved.  Das et al. found an inverse correlation between
ellipticity and relative mass concentration in the centers of
barred galaxies, and noted that such a correlation is predicted by
simulations which show a progression in this direction during
nuclear gas accretion.  However, there was no evidence that this
progression has gone so far in any real galaxy to have dissolved
its bar, as it does in simulations.  One wonders if the balance
between viscous accretion and star formation in galaxy simulations
is a fair representation of the balance in real galaxies, where
the interstellar medium is far more complicated with supersonic
turbulence and magnetic fields.

The uncertainty of the present observations regarding bar
dissolution involves sampling biases. The small sizes and high
restframe surface brightnesses of our barred galaxies suggest they
may be only a subsample of high-z barred galaxies. If the barred
fraction of the missing (i.e., low surface brightness) galaxies is
the same as in the sample here, then our conclusion about bar
dissolution should hold. But if the missing galaxies have a higher
bar fraction than the local value, then some bar dissolution might
have occurred over a Hubble time.

Figures \ref{fig:barvsz}, \ref{fig:8oddbars} and
\ref{fig:radial147310} suggest there are a significant number of
bar-like structures out to at least $z\sim1$ that are very clumpy
or located at non-central places in very clumpy disks. These
morphologies suggest that some bars form in gas-rich disks by
dynamical instabilities in the gas (Mayer \& Wadsley 2004), and
that these instabilities may also trigger a large amount of star
formation (the blue colors of the large clumps in chain galaxies
and clump-clusters were discussed in Papers I and II). The
look-back times in Figure \ref{fig:barfractz} are typically half
the Hubble time for these cases, which implies that some bars
formed late.  If the stars in these bars formed at the same time,
then the present day versions of these galaxies should have
somewhat young-looking bars ($\leq7$ Gyr old).  In addition, if
these bars dissolve and form bulges, then the bulges should
contain relatively young stars too.

\section{Conclusions}
I-band images of the deep field of the Tadpole galaxy suggest the
fraction of barred galaxies is approximately constant,
$0.23\pm0.03$ out to $z\sim1.1$. Beyond this, we cannot observe
bars easily because of inadequate angular resolution and
rest-frame color shifts, but galaxies with inner twisted isophotes
continue out to $z=1.84$. The observed bar fraction decreases with
increasing galaxy inclination, as it does for local galaxies, and
it decreases for smaller angular sizes as the bar lengths become
comparable to the resolution. Correcting for these losses, we
estimate that the bar fraction in the Tadpole deep field is
$\sim0.4$ out to $z\sim1.1$, and we point out that this fraction
is comparable to the local bar fraction, depending on Hubble type.
If the bar fraction is indeed constant with time, and if a
significant fraction of bars form during galaxy interactions
(which were more frequent in the past), then most bars do not
dissolve in a Hubble time following secular evolution.

The physical sizes and exponential scale lengths in the Tadpole
field barred galaxies are comparable to the sizes of local type
SBm galaxies.  Cosmological dimming is probably not affecting this
result because the restframe surface brightnesses of the distant
galaxies are larger than for local galaxies, and because the
number of scale lengths in a galaxy major axis is about the same
as in local galaxies.  If the distant galaxies grow into today's
galaxies, then this growth has to be gentle enough to avoid
disrupting the disks.

Eleven examples of bar-like structures were found in clump-cluster
galaxies, which are probably face-on versions of chain galaxies
(Paper II).  These systems are not present in the same form in the
modern Universe and may represent an early stage in disk galaxy
formation and bar formation. If so, then some bars would seem to
be able to form by large-scale instabilities in primarily gaseous
disks.  Other odd cases included very clumpy bars in disk galaxies
that have approximately exponential radial profiles. These could
also be examples where bars are forming in the presence of a
significant amount of gas, producing a few very bright
star-formation clumps as an intermediate step.

Acknowledgments: We are very grateful to J. Blakeslee, H. Ford,
and J. Mack of the ACS team for providing up-to-date fully reduced
combined images in advance of public release. Helpful comments by
the referee are appreciated. The contribution by B.G.E. was
partially funded by NSF Grant AST-0205097.

\begin{deluxetable}{lccccccccc}
\tablewidth{0pt} \tablecaption{Barred Galaxy Identifications}
\tablehead{\colhead{ID}
& \colhead{Bar}
& \colhead{z}
& \colhead{Diam.}
& \colhead{Diam.}
& \colhead{Scale Length}
& \colhead{Scale Length}
& \colhead{W/L}
& \colhead{$m_{I}$}
& \colhead{$M_I$}\\
\colhead{}
& \colhead{Type}
& \colhead{}
& \colhead{(arcsec)}
& \colhead{(kpc)}
& \colhead{(arcsec)}
& \colhead{(kpc)}
& \colhead{}
& \colhead{}
& \colhead{}
} \startdata
   102 & B &       1.22 &       2.45 &     20.6 &    0.49 &  4.1  &  0.81 &       21.8 &      -22.9\\
   127 & T &        0.66 &       2.68 &     18.8 &   0.30 &  2.1  &     0.72 &       21.2 &      -21.9\\
   234 & B &        0.14 &       6.08 &     15.7 &   0.67 &  1.7  &     0.70 &       20.0 &      -19.2\\
   241 & T &        0.38 &       3.97 &     20.9 &   0.44 &  2.3  &     0.53 &       19.7 &      -21.9\\
   278 & B &        0.43 &       7.56 &     42.8 &   0.62 &  3.5  &     0.94 &       18.6 &      -23.3\\
   396 & T &        0.76 &       1.70 &     12.6 &   0.68 &  5.1  &     0.72 &       22.6 &      -20.8\\
   431 & T &        0.19 &       2.24 &      7.3 &   0.79 &  2.6  &     0.60 &       23.0 &      -17.0\\
   432 & B &        0.81 &       1.95 &     14.8 &   0.32 &  2.5  &     0.70 &       22.6 &      -21.0\\
   448 & B &        0.33 &       1.91 &      9.2 &   0.33 &  1.6  &     0.50 &       24.5 &      -16.8\\
   700 & B &        0.44 &       2.83 &     16.3 &   0.30 &  1.8  &     0.58 &       22.4 &      -19.6\\
   770 & B &       1.09 &       1.46 &     12.0 &    0.19 &  1.5  &    0.72 &       22.7 &      -21.7\\
   828 & B &        0.57 &       1.91 &     12.5 &   0.26 &  1.7  &     0.43 &       23.2 &      -19.4\\
   848 & B &        0.28 &       1.84 &      7.9 &   0.37 &  1.6  &     0.90 &       22.6 &      -18.3\\
  1146 & B &        0.78 &       2.33 &     17.5 &   0.42 &  3.2  &     0.64 &       22.4 &      -21.0\\
  1245 & T &        0.35 &       1.29 &      6.5 &   0.18 &  0.9  &     0.79 &       23.5 &      -17.9\\
  1247 & T &        0.92 &       1.42 &     11.2 &   0.27 &  2.1  &     0.38 &       24.4 &      -19.5\\
  1421 & B &        0.83 &       2.55 &     19.5 &   0.51 &  3.9  &     0.84 &       22.6 &      -21.1\\
  1904 & T &        0.88 &       1.30 &     10.2 &   0.24 &  1.9  &     0.48 &       24.3 &      -19.5\\
  2155 & T &        0.88 &       1.81 &     14.1 &   0.24 &  1.9  &     0.68 &       22.4 &      -21.4\\
  2555 & T &        0.25 &        .99 &      3.9 &   0.20 &  0.8  &     0.90 &       23.6 &      -17.0\\
  2588 & B &       1.11 &       1.38 &     11.4 &    0.24 &  2.0  &    0.60 &       23.2 &      -21.2\\
  2694 & B &        0.07 &       4.83 &      7.2 &   0.70 &  1.0  &     0.63 &       20.1 &      -17.7\\
  2810 & T &        0.84 &       1.39 &     10.7 &   0.20 &  1.5  &     0.67 &       23.6 &      -20.0\\
  2845 & B &        0.69 &       1.85 &     13.3 &   0.31 &  2.2  &     0.98 &       21.7 &      -21.4\\
  3483 & T &        0.26 &       1.24 &      5.1 &   0.24 &  1.0  &     0.65 &       23.9 &      -16.7\\
  3555 & T &       1.84 &        .75 &      6.4 &    0.16 &  1.3  &    0.74 &       24.7 &      -21.1\\
  3983 & T &        0.71 &       1.54 &     11.2 &   0.30 &  2.2  &     0.85 &       23.2 &      -20.0\\
  4129 & B &        0.24 &       3.14 &     12.2 &   0.52 &  2.0  &     0.65 &       21.7 &      -18.8\\
  4227 & B &        0.80 &       3.70 &     28.0 &   0.32 &  2.4  &     0.66 &       21.4 &      -22.1\\
  4286 & B &        0.83 &       2.27 &     17.4 &   0.24 &  1.8  &     0.90 &       22.0 &      -21.6\\
  4451 & T &        0.44 &        .60 &      3.4 &   0.25 &  1.4  &     0.70 &       24.8 &      -17.2\\
  4777 & T &       1.15 &       1.19 &      9.9 &    0.19 &  1.6  &    0.66 &       23.7 &      -20.9\\
  4818 & T &        0.75 &        .59 &      4.4 &   0.27 &  2.0  &     0.84 &       24.4 &      -19.0\\
  4918 & B &        0.80 &       1.12 &      8.4 &   0.15 &  1.1  &     0.62 &       24.1 &      -19.5\\
  4967 & T &        0.77 &       1.76 &     13.2 &   0.34 &  2.6  &     0.73 &       22.4 &      -21.1\\
  5001 & T &       1.09 &       1.21 &     10.0 &    0.23 &  1.9  &    0.68 &       23.5 &      -20.9\\
  5182 & T &        0.75 &       2.80 &     20.7 &   0.38 &  2.8  &     0.28 &       23.1 &      -20.2\\
  5286 & B &        0.76 &       4.00 &     29.7 &   0.57 &  4.3  &     0.63 &       21.6 &      -21.8\\
  S1 & B &           x &         3.00 &      x &     0.72 &  x &     0.74 &       21.8 & x\\
  S2 & B &           x &         3.95 &      x &     0.78 &  x &      0.57 &       21.2 & x\\
  S3 & B &           x &         5.46 &      x &     0.45 &  x &      0.38 &       20.6 & x\\
  S4 & T &           x &         3.12 &      x &     0.57 &  x &     0.38 &       21.3 & x\\
  S5 & T &           x &         3.25 &      x &     0.41 &  x &     0.89 &       20.5 & x

\enddata
\tablecomments{ID numbers and redshifts are from Benitez et al.
(2004). Bar type B is for a clearly barred galaxy, T is for a
galaxy without a clear bar but with an inner isophotal twist
indicative of a possible bar. Diameters in kpc are from Eq. 2.
$W/L$ is the isophotal width to length ratio. $m_I$ is the AB
isophotal magnitude in the F814W filter. Galaxies S1-5 are spiral
galaxies not cataloged by Benitez et al. because they are close to
the Tadpole galaxy or bright stars. The $x$'s indicate there are
no redshifts or redshift-derived properties for these galaxies.
Their coordinates (J2000) are: S1 = $16^{\rm h}06^{\rm
m}17.3443^{\rm s}$ $55^\circ26^\prime30.283^{\prime\prime}$, S2 =
$16^{\rm h}06^{\rm m}13.7069^{\rm s}$
$55^\circ26^\prime47.627^{\prime\prime}$, S3 = $16^{\rm h}06^{\rm
m}21.7162^{\rm s}$ $55^\circ26^\prime07.421^{\prime\prime}$, S4 =
$16^{\rm h}06^{\rm m}10.4523^{\rm s}$
$55^\circ26^\prime27.142^{\prime\prime}$, S5 = $16^{\rm h}06^{\rm
m}05.5798^{\rm s}$ $55^\circ26^\prime16.090^{\prime\prime}$.}
\end{deluxetable}

\begin{deluxetable}{lcccccc}
\tablewidth{0pt} \tablecaption{Barred clump-cluster galaxies}
\tablehead{ \colhead{ID} & \colhead{z} &
\colhead{Diameter} & \colhead{Diameter} &
\colhead{W/L} & \colhead{$m_{I}$} & \colhead{$M_I$}\\
\colhead{}
& \colhead{}
& \colhead{(arcsec)}
& \colhead{(kpc)}
& \colhead{}
& \colhead{}
& \colhead{}

} \startdata
    91 &        0.32 &       1.22 &      5.8 &        0.85 &       22.8 &      -18.4\\
   689 &        0.98 &       3.83 &     30.8 &        0.51 &       21.2 &      -22.9\\
   971 &        0.84 &       1.28 &      9.9 &        0.87 &       23.2 &      -20.5\\
  1253 &       1.14 &       1.42 &     11.8 &        0.84 &       22.9 &      -21.6\\
  2861 &        0.70 &       1.41 &     10.2 &        0.80 &       23.1 &      -20.1\\
  3168 &        0.57 &       1.20 &      7.9 &        0.70 &       24.0 &      -18.7\\
  3429 &       1.30 &       1.69 &     14.3 &        0.53 &       23.3 &      -21.6\\
  3780 &       1.11 &       1.37 &     11.3 &        0.86 &       23.7 &      -20.8\\
  4679 &        0.80 &       2.98 &     22.5 &        0.34 &       22.6 &      -20.9\\
  4903 &        0.08 &       1.34 &      2.2 &        0.82 &       23.1 &      -15.0\\
  4935 &        0.75 &       1.89 &     14.0 &        0.49 &       23.5 &      -19.8

\enddata
\tablecomments{ID numbers and redshifts are from Benitez et al.
(2004). $m_I$ is the AB isophotal magnitude in the F814W filter.}

\end{deluxetable}

\begin{figure}
\plotone{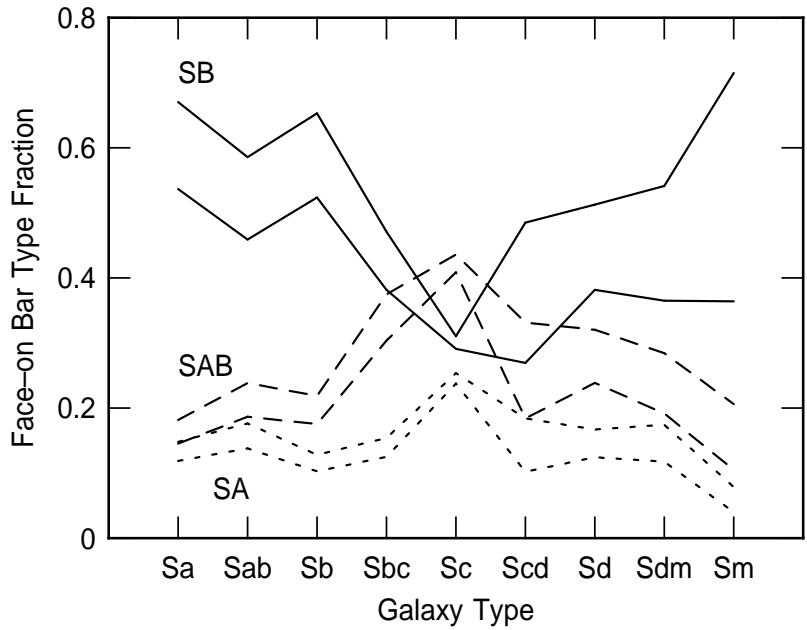} \caption{Fractions of galaxies in the RC3 with
$W/L>0.5$ that have bar types SB, SAB, or SA (non-barred) as a
function of the spiral type. The fraction totals 1 within each
galaxy type (i.e., vertically). The upper line for each category
is the fraction relative to the galaxies with known bar types,
while the lower line is the fraction relative to all galaxies with
RC3 T types, including those with unknown bar
types.}\label{fig:barfracttype}
\end{figure}

\end{document}